NOËLLE CARBONELL, SUZANNE KIEFFER

# DO ORAL MESSAGES HELP VISUAL SEARCH?

**Abstract.** A preliminary experimental study is presented, that aims at eliciting the contribution of oral messages to facilitating visual search tasks on crowded visual displays.
Results of quantitative and qualitative analyses suggest that appropriate verbal messages can improve both target selection time and accuracy. In particular, multimodal messages including a visual presentation of the isolated target together with absolute spatial oral information on its location in the displayed scene seem most effective. These messages also got top-ranking ratings from most subjects.

## 1. CONTEXT AND MOTIVATION

### 1.1. Multimodality: State of the Art

Numerous forms of speech-based input multimodality have been proposed, implemented and tested. Combinations of speech with gestural modalities have been studied extensively, especially combinations of speech with modalities exploiting new input media, such as touch screens, pens, data gloves, haptic devices. Both usability and implementation issues have been considered; see, among others, (Oviatt, De Angeli and Kuhn, 1997) [1] or (Robbe, Carbonell and Dauchy, 2000) [2] for the first category of issues, (Nigay and Coutaz, 1993) for the second category.

Contrastingly, speech combined with text and graphics has only motivated a few studies. As an output modality, speech is mostly used either as a substitute for standard visual presentation modes (*cf.* phone services) or for supplementing deficiencies in visual exchange channels. Recent research efforts have been focusing on two main application domains:

– Providing blind or partially sighted users with easy computer access (*e.g.*, Grabowski and Barner, 1998; Yu, Ramloll and Brewster, 2001), and
– Implementing appropriate interaction facilities in contexts of use where access to standard screen displays is difficult or even impossible. This is the case, for instance, in contexts where only small displays are available (*e.g.*, PDAs and wearable computers), or in situations where the user's gaze is involved in other activities (*e.g.*, while driving a car); see, for instance, (Baber, 2001) concerning the first class of situations, and (De Vries and Johnson, 1997) concerning the second one.

---

[1] On speech and pen.

[2] On speech and finger gestures on a touch screen.

However, there is not yet, at least to our knowledge, a substantial amount of scientific work on the integration of speech into the output modalities of standard user interfaces, that is interfaces intended for standard categories of users using standard application software in standard environments and contexts of use; graphical user interfaces (or GUIs) implementing direct manipulation design principles (Shneiderman, 1983) are still prevailing.

Published research on output forms of multimodality including speech amounts to usability studies of the role of oral comments in multimedia presentations, such as (Faraday and Sutcliffe, 1997), and contributions to the automatic generation of multimedia presentations (*cf.* André and Rist, 1993; Maybury, 2001).

This lack of interest for output forms of multimodality on the part of the research community may result, at least partly, from the fact that, although *multimedia* and *multimodality* refer to different concepts, these terms are often used as synonyms, especially when applied to system outputs. Precise definitions are presented in the next paragraph.

*1.2. Definitions: Multimodality versus Multimedia*

(Coutaz and Caelen, 1991), (Maybury, 1993; 2001) and (Bernsen, 1994), among others, define '*media*' and '*modalities*' contrastingly.

They use the first term for referring to the hardware and software channels through which information is conveyed, and the second one for designating the coupling of a medium with the interpretation processes required for transforming physical representations of information into meaningful symbols or messages. In other words, and focusing on output media and modalities:

> '… by *media* we mean the carrier of information such as text, graphics, audio, or video. Broadly, we include any necessary physical interactive device (*e.g.*, keyboard, mouse, microphone, speaker, screen). In contrast, by *mode* or *modality* we refer to the human senses (more generally agent senses) employed to process incoming information, *e.g.*, vision, audition, and haptics.'
> (Maybury, 2001)

To characterize the various possible combinations of modalities, taxonomies have been proposed. In (Coutaz and Caelen, 1991), multimodality is characterized in terms of the strategies available for coordinating the use of different modalities temporally and semantically.

In addition, (Coutaz *et al.*, 1995) defines four properties which prove useful for comparing modalities in terms of their expressive power (*i.e.*, complementarity versus equivalence), and for defining their semantic contribution when used in multimodal contexts (*i.e.*, redundancy versus complementarity).

As for Bernsen's taxonomy (*cf.* Bernsen, 1994), it is a thorough inventory of the output modalities available to user interface designers.

*1.3. Motivation and Objectives*

The definition of the information content and semantics of multimedia presentations is commonly viewed as the responsibility of experts in the specific application domain considered. It is seldom viewed as lying within the scope of research on human-computer interaction. Therefore, the frequent assimilation of multimodality to multimedia may explain why the design of appropriate multimodal system responses has raised but little interest in the user interface research community, especially from an ergonomic angle, save for studies focused on specific categories of users or specific contexts of use.

However, if standard users are offered speech facilities together with other input modalities, it is mandatory that the system responses are not limited to visual messages. Communication situations where one interlocutor can speak and the other cannot, are rather unusual. Research is then needed on the usability and software issues concerning the generation of appropriate multimodal system responses in standard human-computer interaction environments, and for standard user categories, including the general public.

The main objective of the preliminary experimental study presented here is to contribute to scientific advances in this research area, in-as-much as it addresses one of the major usability issues relating to the generation of effective oral system messages, namely:

> How to design oral messages which facilitate the visual exploration of crowded displays?
>
> In particular, how to design messages which effectively help users to locate specific graphical objects on such displays?

Resorting to deictics and visual enhancements of graphical targets is a solution which seems "natural". However, it is no more effective than the sole visual enhancement of the target.

Another approach is to implement a human-like animated embodiment of the system, to visualise it on the screen and to endow it with a pointing device; see, for instance the PPP [3] persona which impersonates a car dealer, and uses a pointing stick for attracting the user's attention on the assets of the currently displayed car (André, 1997). However, the contribution of personae to the usability and effectiveness of human-computer interaction is still unclear (*cf.* Mulken, André and Müller, 1999). Further testing is required in order to determine the usefulness of visual system embodiments in graphical user interfaces.

These reasons explain why we chose to focus first on assessing whether oral messages including spatial information actually facilitate the visual exploration of complex displays, especially the localization of graphical targets, in the context of standard human-computer interaction and environment.

---

[3] PPP means 'Personalized Plan Based Presenter'. See also Candace Sidner's robot, a penguin which can point at locations on large horizontal displays with its beak (Sidner and Dzikovska, 2002).

We selected visual search as the experimental task for the following reasons. It is one of the few human visual activities, besides reading, that have motivated a significant amount of psychological research (*cf.* Henderson and Hollingworht, 1998; Findlay and Gilchrist, 1998; Chelazzi, 1999). The design of numerous computer applications may benefit from a better knowledge of this activity, especially applications for the general public such as:

- Online help to current interactive application software. For instance, novice users interacting with present graphical interfaces are often confused by the increasing number of toolbars and icons displayed concurrently.
- Map reading environments (*cf.* geographical applications), and navigation systems in vehicles.
- Data mining in visualisations of very large data sets; see, for instance, the complex hyperbolic graph visualisations proposed in (Lamping, Rao and Pirolli, 1995), or the treemap representations in (Fekete and Plaisant, 2002), and (Shneiderman, 1996; Card and Mackinlay, 1997; Card, Mackinlay and Shneiderman, 1999) for a general overview of visualisation techniques and their use.

The methodology and experimental set-up of the experiment presented here are described in the next section, together with the underlying scientific hypotheses. Then, quantitative and qualitative results are presented and discussed in section 3. Future research directions stemming from these results are described in the general conclusion (*cf.* section 4).

## 2. METHODOLOGY AND EXPERIMENTAL SET-UP

*2.1. Overall experimental protocol*

To assess the potential contribution of oral spatial information to facilitating visual search, we designed a preliminary experiment with:
- target presentation mode as *independent* variable,
- target search+selection time and target selection accuracy, as *dependent* variables.

Eighteen subjects were to locate and select visual targets in thirty six complex displayed scenes, using the mouse. They were requested to carry out target localization and selection as fast as they could. Colour displays only were used.

Each scene display was preceded by one out of three possible presentations of the corresponding target:
- Display of the isolated target at the centre of the screen during three seconds (visual presentation or *VP*);

- Oral characterization of the target (*i.e.*, name of the relevant graphical object), plus spatial information on its position in the scene (oral presentation, *OP*);
- Simultaneous display of the visual and oral presentations of the target (*i.e.*, multimodal presentation, *MP*).

These three sets of thirty six dual stimuli [4] defined three experimental situations, namely the VP, OP and MP conditions. In the MP condition, the visual and oral presentations used in the VP and OP conditions respectively were presented simultaneously.

Subjects were randomly split up into three groups, so that each subject processed twelve pairs of stimuli *per* condition, and each pair of stimuli was processed by six subjects.

In order to neutralize possible task learning effects, the processing order was counterbalanced inside each group of subjects as follows: VP-PO-MP (three subjects), and OP-VP-MP (three subjects). All subjects performed the MP condition last.

The size of groups in usability studies or cognitive ergonomics experimental studies seldom exceeds six subjects, most likely because analysing the behaviours of subjects performing realistic tasks in realistic environments is indeed a costly undertaking; for instance, (Ahlberg, Williamson and Shneiderman, 1995) reports an experimental evaluation of three different user interfaces (meant for exploring information spaces) which also involved eighteen subjects split up into three groups of six subjects each, one group *per* user interface.

The overall set-up is summed up in table 1.

*Table 1: Overall task set-up.*

*$G_i$: group of 3 subjects (3 x 6 = 18 subjects)*
*$P_i$: set of 12 visual scenes (3 x 12 = 36 scenes)*

| Group | VP | OP | MP | Group | OP | VP | MP |
|---|---|---|---|---|---|---|---|
| G1 | P1 | P2 | P3 | G4 | P2 | P1 | P3 |
| G2 | P3 | P1 | P2 | G5 | P1 | P3 | P2 |
| G3 | P2 | P3 | P1 | G6 | P3 | P2 | P1 |

Experimental design choices were mainly motivated by the intent to assess the soundness of the three following working hypotheses which use the VP condition as the reference situation:

A. Multimodal presentations of targets will reduce selection times and error rates in comparison with visual presentations.

---

[4] That is 108 pairs of stimuli (36 scenes x 3 presentation modes), each pair consisting in a presentation (visual, oral or multimodal) of a target followed by the display of the scene including this target.

B. Oral presentations of targets will also improve accuracy, compared to visual presentations.
C. The type of spatial information included in oral target presentations of will influence selection times and error rates. In particular, absolute and relative spatial indications will prove more effective than references to *a priori* knowledge (*cf.* subsection 2.3), and absolute spatial information will be more effective than relative spatial indications.

These hypotheses are based on common sense reasoning, in the absence, at least to our knowledge, of earlier published results and models stemming from experiments comparable to ours, that is involving similar tasks and interaction environments.

Targets were unique graphical objects in the scenes including them. In addition, oral messages were designed so as to designate targets unambiguously. However, since they were presented out of context, they might be easily confounded with other graphical objects in the scene during visual search for the target. Then, unambiguous linguistic designations of targets being likely to prevent selection errors due to possible visual confusions, it follows that accuracy would be higher in the OP and MP conditions than in the VP condition.

As regards selection times, we assumed that subjects would use the spatial information included in oral messages, and that this information would enable them to focus visual search for the current target on a rather limited area in the scene. Scenes being complex and displayed on a rather large screen [5], we should then observe sensibly shorter selection times in the OP and MP conditions than in the VP condition.

Concerning hypothesis C, absolute spatial information implies a one-step target localization process, while relative spatial information induces a two-step visual search; the latter is then less effective than the former as regards selection time, cognitive workload and accuracy. As for references to *a priori* knowledge, they involve more complex cognitive processes than the two previous types of spatial information; in addition, cultural knowledge varies greatly among users.

In the remainder of the section, further information is given on:
− the criteria used for selecting visual scenes and targets (2.2);
− the structure and information content of oral messages (2.3);
− subjects' profiles (2.4);
− the experimental set-up (2.5);
− the methodology adopted for analysing subjects' results (2.6).

---

[5] Scenes included several scores of graphical objects and were displayed on a 21 inches screen.

*2.2. Scene Selection Criteria*

Most visual scenes were taken from currently available Web pages in order to provide subjects with realistic task environments.

They were classified according to criteria stemming from Bernsen's taxonomy of output modalities (*cf.* Bernsen, 1994), our aim being to investigate the possible influence of the type of visual scenes displayed on target selection times and accuracy.

Our classification was derived from the graphical categories in Bernsen's taxonomy as follows. We focused on static graphical displays exclusively [6], on the ground that the localization and selection of moving targets in animated visual presentations is a much more complex activity than the selection of still targets in static visual presentations. Issues relating to the exploration of visual animated scenes will be addressed at a later stage in our research.

We established two main classes of static presentations:
- *Class 1* comprises displays of structured or unstructured collections of symbolic or arbitrary graphical objects, such as maps, flags, graphs, geometrical forms (*cf.* classes 9, 11, 21, 25 in Bernsen's taxonomy);
- *Class 2* includes displays of realistic objects or scenes, namely photographs or naturalistic drawings figuring complex real objects (*e.g.*, monuments) or everyday life environments, such as views of rooms, town or country landscapes, … (*cf.* class 10 in Bernsen's taxonomy).

Half of the thirty six visual scenes belonged to class 1, and the other half to class 2. Class 1 and class 2 scenes in each of the three subsets described in subsection 2.1 (*cf.* table 1) were randomly ordered.

Targets were objects or component parts of complex objects (*cf.* the complex real objects in class 2).

They were chosen according to the following criteria. An acceptable target was a unique definite graphical object that could be designated unequivocally by a short simple verbal phrase. Although all targets were unique, some of them could be easily confused (visually) with other objects in the scene.

In order to avoid task learning effects, target visual properties [7] and position were varied from one scene to another.

---

[6] *cf.* the five types of static graphical presentations in Bernsen's taxonomy, namely classes 9, 10, 11, 21, 25.

[7] These properties mainly include: colour, shape, orientation and size (within the limits of the fixed size target presentation box).

*2.3. Message structure and content*

All messages included a noun phrase meant to designate the target unequivocally. For instance, "the pear" refers to the target unequivocally in the realistic scene reproduced in figure 1.

For any target, we chose the shortest simplest noun phrase that could characterize it without ambiguity and redundancy. For instance, one of the scenes represented geometrical figures and included several squares. The target we chose was the smallest square and the only pink one. We referred to one of these features, the colour, to appropriately reduce the scope of the substantive "square" in the oral message; "the pink square" is a noun phrase that refers to the target without ambiguity or redundancy. On the other hand, the use of the substantive "pear" is sufficient for referring unambiguously to the target in the scene reproduced in figure 1.

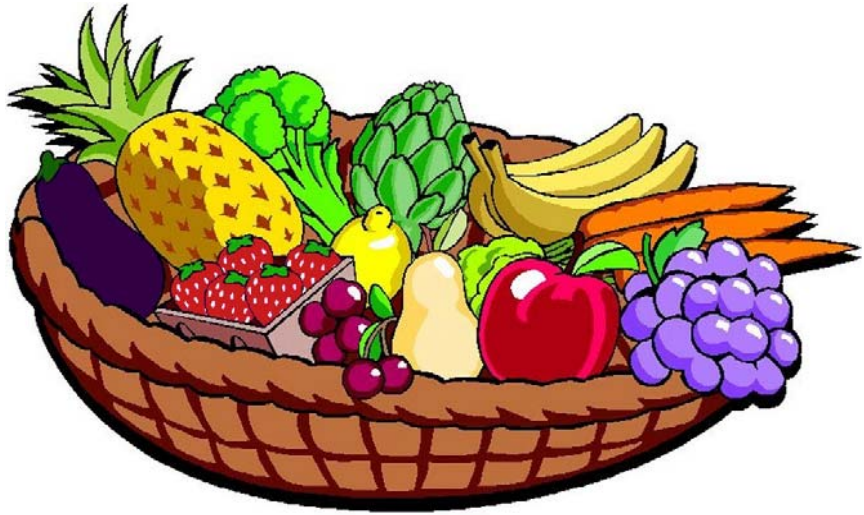

*Figure 1: Basket with fruit (class 2 picture).*
*Oral message: "On the left of the apple, the pear."*

We experimented three types of spatial information in the verbal phrases referring to target locations, using an *ad hoc* classification inspired by the taxonomy presented in (Franck, 1998):
- Absolute spatial information (*ASI*), such as "on the left/right" or "at the top/bottom";
- Relative spatial information (*RSI*), for instance "on the left of the apple" (*cf.* figure 1);

- Implicit spatial information (*ISI*), that is spatial information that can be easily inferred from common *a priori* knowledge and the visual context; for instance, it is easy to locate and identify the Mexican flag among twenty other national flags from the simple message "The Mexican flag.", if the scene represents a planisphere, and national flags are placed inside the matching countries (*cf.* figure 2).

Messages included one or two spatial phrases, according to scene complexity; one or two types of spatial information, namely ASI+RSI or ASI+ISI, were grouped in phrase pairs.

Careful attention was paid to the choice of spatial prepositions (Burhans, Chopra and Srihari, 1995).

In order to make the assessment of hypothesis C possible, all messages had the same syntactical structure, in order that information content was the only factor pertaining to the design of messages that could influence localization times and selection errors. The following structure, which emphasizes spatial information, was adopted for most messages, some ISI messages including no spatial phrase:

*[Spatial_phrase] + Noun_phrase (designation)*

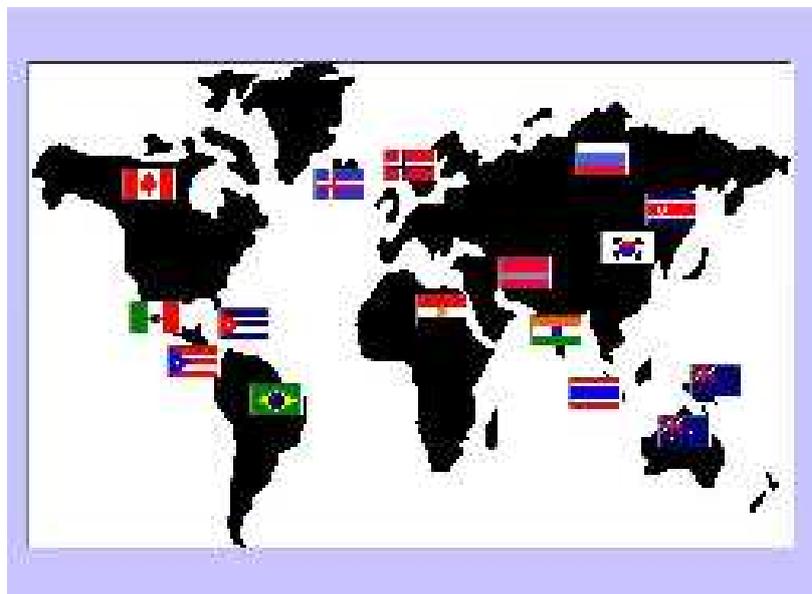

*Figure 2: National flags (class 1).*
*Oral message: "The Mexican flag."*

This structure was preferred to the usual one (Noun_phrase+Spatial_phrase), based on the assumption that information in messages would be more effective if it was presented in the same order as it would be processed by the user.

*2.4. Subjects' Profiles*

As this study involved a restricted number of subjects (18) and was a first attempt at validating hypotheses A, B and C, we defined strong constraints on subjects' profiles in order to reduce inter-individual diversity, especially regarding task performance, and limit the number of factors that might influence subjects' performances.

To achieve homogeneity, we selected 18 undergraduate or graduate students in computer science with normal eyesight [8] and ages ranging from 22 to 29. Thus, all participants were expert mouse users with alike quick motor reactions, familiar with visual search tasks, and capable of performing the chosen experimental tasks accurately and rapidly.

*2.5. Experimental Set-up*

First, the experimenter presented the overall experimental set-up. Then, after a short training (6 target selections in the VP situation), each subject processed 12 scenes *per* situation, in the order VP+OP+MP or OP+VP+MP.

Before each change of condition (*i.e.*, before each change in target presentation), the experimenter explained the new specific set-up to the subject.

For each visual scene:
- The target was first presented, during three seconds:
  − either visually in a fixed-size box in the centre of the screen,
  − or orally (together with a blank screen),
  − or orally and visually, simultaneously.
- Then, a button appeared in the centre of the screen together with a written message requesting the subject to click on the button for launching the display of the scene. Therefore, at the beginning of each target selection step, the mouse was positioned in the centre of the screen, making it possible to compare subjects' selection times.
- The next target was presented as soon as the subject had clicked on any object in the current displayed scene.

At the end of the session, subjects had to fill in a questionnaire requesting them to rate the difficulty of each task using a four degree scale (ranging from "very easy" to "very difficult"). The session ended up in a debriefing interview.

---

[8] save for one subject who was slightly colour-blind.

*2.6. Analysis methodology*

Quantitative results comprise:
- average {localization + selection} times, and
- error (*i.e.,* wrong target selections) counts or percentages,

computed over all subjects and scenes, as well as *per* class of scenes and *per* type of oral message.

We tested the statistical significance of these quantitative results whenever a sufficient number of samples was available.

Qualitative analyses of subjects' performances, especially comparisons between the numbers of target selection errors in the VP, OP and MP conditions, provided useful information for defining further research directions.

In order to elicit the possible factors at the origin of selection errors, scenes and targets were characterized using the following features:
- Scene characterization:
  - complexity (according to the number of displayed objects); and,
  - for class 1 scenes only, visual structure (*e.g.*, random layout of objects; tree, crown or matrix structures; …).
- Target characterization:
  - position on the screen (centre, left, …);
  - visual salience;
  - familiarity versus unfamiliarity (oddness);
  - ambiguity (*i.e.*, the number of possible visual confusions with other graphical objects in the scene).

Quantitative and qualitative results are presented and discussed in the next section.

## 3. RESULTS: PRESENTATION AND DISCUSSION

*3.1. Quantitative Results*

Results were computed over 34 scenes; two scenes (both in class 1) had to be excluded from both quantitative and qualitative analyses by reason of technical incidents. Therefore, the corpus of experimental data we actually used comprised the results of 612 visual search tasks performed by 6 different subjects, that is 204 *per* condition (VP, OP, or MP).

Statistical tests, mainly t-tests, were performed on these performance data (successes/failures and execution times), especially on the three sets of data collected in the three conditions.

*3.1.1. Global Analysis*

*Presentation.* Concerning the order of conditions (VP then OP, versus OP then VP), comparisons between subjects' performances in each of the two groups yielded no significant inter-group differences. These results indicate that no perceptible task learning effect occurred in the course of the experiment. The absence of any such effect is not surprising, due to the low number of tasks in each condition (i.e., 12) and the brief duration of the overall session (about 10 min).

*Table 2: Results per target presentation mode.*
*Upper half: best results are in bold type, and lowest ones are underscored.*
*Lower half: significant statistical results are in bold type.*

| Target presentation mode | Number of errors | Average selection time (sec.) | Standard deviation (sec.) |
|---|---|---|---|
| VP | <u>31</u> | 2.83 | 1.70 |
| OP | 14 | <u>3.92</u> | <u>3.50</u> |
| MP | **8** | **2.70** | 1.93 |

| Target presentation mode | Number of errors | | Average selection time (sec.) | |
|---|---|---|---|---|
| VP versus OP | t= -2.70 | **p=0.007** | t=+3.79 | **p=0.0002** |
| VP versus MP | t=-3.94 | **p<0.0001** | t=-0.70 | p=0.4852 |
| OP versus MP | t=-1.31 | p=0.189 | t=-4.20 | **p<0.0001** |

Therefore, learning effects may be rightly excluded from the factors to be considered in the interpretation of the quantitative and qualitative results of our analyses, even as regards the MP condition although it was performed last by all subjects.

As for selection accuracy, oral messages proved much more effective than visual target presentations, as shown by comparisons between the VP and OP conditions (*cf.* table 2).

The total number of errors in the OP condition decreased by 55%. However, selection was slower in the absence of prior visualizations of isolated targets, that is in comparison with the VP and MP conditions. Average selection time in the OP condition increased by over 38% compared to the OP condition.

These differences are statistically significant.

Table 2 also shows that multimodal presentations of targets reduced both selection times compared to oral presentations, and error rates in comparison with visual presentations, both results being statistically significant. These results are in keeping with the previous ones.

*Interpretation and discussion.* As regards accuracy, spatial indications and target verbal designations included in messages may have reduced the frequency of visual confusions, the former by reducing the scope of the visual search for the target, the latter by preventing confusions between the target and graphical objects of similar visual appearance in the scene.

Significantly lower average selection time in the OP condition, together with a much higher standard deviation, may be explained by the fact that subjects were unfamiliar with the visual search tasks they had to carry out in the OP condition; this situation being rather unusual compared to the VP and MP situations which occur frequently in everyday life. Therefore, the higher variability of selection times in the OP condition may be assumed to reflect the high inter-individual diversity of cognitive abilities and processes.

However, the slower selection times observed in the OP condition may be explained more satisfactorily by the intrinsic differences between the tasks subjects had to perform in the OP and the VP conditions. Visual search of a visually known graphical target is a much less complex task than searching for a graphical object that matches a given verbal specification, even if only a part of the scene needs to be explored thanks to spatial verbal information.

(Bieger and Glock, 1986) observed similar effects on the performances of subjects in an experiment that aimed at comparing the efficiency of text and graphics for presenting various types of information in instructional material. In particular, concerning spatial information [9], Bieger and Glock found that subjects who were to use textual presentations of spatial information made fewer task execution errors than those who were given graphical presentations of the same information; on the other hand, the latter completed the prescribed tasks faster.

However, further comparisons between this study and ours would be meaningless, due to important design differences between the two experimental protocols: in Bieger's and Glock's study, the tasks subjects had to carry out were procedural assembly tasks (versus target selection in our experiment), and the modalities considered were text and graphics (versus speech and graphics).

Concerning subjects' performances in the MP condition, comparisons with their performances in the other conditions suggest that, in this condition, they succeeded in making the most of the information provided by each modality, thus compensating for the weaknesses of each unimodal presentation mode. A likely interpretation is that they achieved:

– easy disambiguation of possible confusions between targets and other objects of similar appearance in the scene, thanks to spatial and denominative verbal information; and
– rapid identification of targets thanks to visual information, matching based on visual characteristics being faster than identification based on abstract

---

[9] Namely the final position and orientation of elements in the workspace (assembly tasks).

properties or stereotyped mental representations, which involves complex decision-making processes.

Surprisingly enough, spatial verbal information which, according to our hypotheses, should accelerate target localization, did not actually contribute to reducing selection times significantly, or so it seems.

This finding, together with the slower selection times observed in the OP condition may be explained within the framework of current visual perception models which assume that eye-movements are less influenced by cognitive (top-down) processes than by visual stimuli (bottom-up processes) during visual exploration tasks (Henderson and Hollingworth, 1998). On the other hand, visual exploration in search of a graphical object matching a verbal specification is a task that combines complex, hence slow, cognitive processes with perceptual activity. These differences may explain why slower selection times were observed in the OP condition than in the VP and MP conditions, and why spatial information did not noticeably reduce selection times in the MP condition compared to the VP condition.

This interpretation of the noticeable differences observed between average selection times in the three conditions, fits in with Rasmussen's cognitive model (Rasmussen, 1986), provided that the search of a visually known target is assimilated to a skilled or automatic activity, and the search of a visually unknown target from a verbal specification of its characteristics as a problem-solving activity. The model then predicts that selection time of the known target will be shorter than that of the unknown target, based on the assumption that automatic or skilled responses to stimuli amount to the activation of precompiled schemata or the compilation and activation of pre-defined schemata respectively, while problem-solving involves more complex cognitive processes.

The application of this model to the MP condition suggests the tentative conclusion that subjects used both sources of information optimally within an overall opportunistic strategy favouring visual search over more complex matching processes, matching processes being activated only in cases when visual information is insufficient or ambiguous.

However, although Rasmussen's model constitutes an appropriate framework for predicting subjects' performances, it is too general to provide any meaningful insight into how the strategy underlying these performances is implemented, that is, how multimodal stimuli are processed, how the results of modality specific processes are integrated and unified to produce meaningful coherent interpretations and appropriate actions or motor responses .

Subjects' performances are indeed compatible with cognitive models of multimodal input processing that postulate the existence of high level interactions between unimodal perceptual and interpretative processes rather than early low-level interactions (*cf.*, for instance, Engelkamp, 1992).

However, further research is needed in order to determine which type of interaction (competition, synergism or complementarity) would best account for subjects' results within the framework of our experimental protocol.

For instance, reducing progressively the duration of visual target presentations in the MP condition appears as a promising, but difficult to implement, experimental paradigm for increasing our knowledge of the cognitive processes involved in the processing of multimodal inputs (Massaro, 2002).

*Future work.* To achieve significant advances in the investigation of multimodal perception and interpretation processes, experimental approaches have to overcome major difficulties.

In particular, homogeneous sets of visual search tasks are needed in order to make it possible to achieve meaningful reliable intra- and inter-subject comparisons, as the same scene cannot be processed in the three conditions by the same subject.

The "difficulty" of the visual search tasks proposed to our subjects varied greatly from one scene to another. For instance, in the VP condition, all subjects (6) failed to select the correct target for one scene, and only 8 scene+target pairs occasioned 26 out of the 31 errors observed in this condition. We are currently refining our visual and semiotic characterizations of scenes and targets in order to be capable of defining and generating sets of really equivalent scenes.

The number of participants and scenes should also be increased considerably, so that a greater number of more sophisticated inter-related hypotheses can be tested simultaneously, and their soundness evaluated accurately, thanks to the possible application of appropriate statistical techniques.

*Ergonomic recommendations.* To conclude, the quantitative results presented in this section contribute to validating hypotheses A and B. However, further research is needed to determine whether the significantly longer selection times observed in condition OP are mainly due either to the differences between the tasks subjects performed in the OP condition and those they carried out in the VP and MP conditions, or to the fact that subjects were unfamiliar with the tasks they carried out in this condition.

These results also suggest useful recommendations for improving user interface design.

In order to facilitate visual search tasks in crowded displays without resorting to standard visual enhancement techniques, two novel forms of user support may prove useful alternatives:

a. If accuracy only is sought for, an oral message comprising an unambiguous verbal designation of the graphical target and spatial information on its location in the display will prove sufficient.
b. If both accuracy and rapidity are sought for, a multimodal message will be more appropriate, that is a message comprising a context-free visual presentation of the target together with an oral message including the same information as mentioned in recommendation a.

However, further experimental research is needed to confirm these recommendations beyond doubt, in-as-much as they have been inferred from a relatively small sample of experimental data and measurements.

In addition, oral and multimodal messages should be compared, in terms of effectiveness and comfort, with other forms of user support, such as target visual enhancement through colour, animation, zooming, etc. Until a sufficient amount of experimental data has been collected, recommendations a. and b. should be considered as tentative.

Analyses of results *per* class of scenes and type of messages (*i.e.*, type of spatial information) are presented next. These analyses make it possible to refine and enrich our initial working hypotheses.

*3.1.2 Detailed analysis*

*Results per class of scenes.* Subjects' results, grouped per scene class and target presentation mode, are presented in table 3. Error percentages were computed over:
- 96 samples *per* condition for class 1 (due to the exclusion of two class 1 scenes as explained at the beginning of section 3), and
- 108 samples *per* condition for class 2.

No statistical analysis was performed on these results by reason of the rather small number of samples in each set.

Multimodal messages proved most effective, in comparison with visual and oral presentations, especially for scenes representing symbolic or arbitrary objects. For scenes in class 1, comparisons between the three conditions indicate that errors were reduced in the MP condition by 12.6% (compared to the VP condition) and 5.3% (compared to the OP condition), while average selection times decreased by 0.24 and 1.27 seconds respectively. In short, concerning class 1 scenes, 86% of the selection errors observed in the VP condition did not occur in the MP condition, and selection times were one third longer in the OP condition than in the MP condition.

As for realistic scenes, the average selection time in the MP condition is similar to the VP one (2.40 sec. versus 2.43 sec.) and markedly inferior to the OP one (3.58 sec.), whereas the number of selection errors is similar to the OP one, and inferior to the VP one (by 10.2%).

These results confirm the main hypothesis proposed in the previous subsection for interpreting global results, namely that, in the MP condition, subjects took advantage of both the visual and oral information available, especially for processing class 1 scenes.

For instance, we observed that some subjects lacked the *a priori* knowledge required for taking advantage of the implicit information conveyed by ISI messages; visual information can prove most helpful for achieving successful target identification in such cases. On the other hand, class 1 scenes which consisted in collections of symbolic or arbitrary graphical objects (*e.g.*, flags or geometric figures) favoured visual confusions between targets and similar objects in the displayed collections; verbal designations and spatial information undoubtedly helped subjects to solve possible visual "ambiguities".

*Table 3: Results per target presentation mode and class of scenes.*
*Percentages were computed over the total number of samples (selection tasks) per condition.*

| Target presentation mode | Error percentage | Average selection time (sec.) | Standard deviation (sec.) |
|---|---|---|---|
| VP-C1 | 14.6 | 3.27 | 1.94 |
| VP-C2 | 15.7 | 2.43 | 1.39 |
| OP-C1 | 7.3 | 4.30 | 4.09 |
| OP-C2 | 6.5 | 3.58 | 2.87 |
| MP-C1 | 2 | 3.03 | 2.36 |
| MP-C2 | 5.5 | 2.40 | 1.36 |

The fact that average selection times were consistently longer for class 1 scenes than for class 2 scenes can be explained as follows.

If the target is a familiar object (such as a pan) in a familiar realistic scene (a kitchen, for instance), visual exploration of the scene is facilitated by *a priori* knowledge of the standard structure of such environments and the likely locations of the target object therein. Such knowledge is not available in the case of unrealistic class 1 scenes; the structure of such a scene and the possible locations of the target in it cannot be foreseen using *a priori* knowledge, so that a more careful search, or even an exhaustive exploration, of the scene is necessary for succeeding in locating the target.

This hypothesis may also explain why multimodal target presentations proved most effective for scenes belonging to class 1: both oral and visual information contributed to compensate for the lack of pragmatic *a priori* knowledge.

*Results per type of messages.* Five categories of verbal messages were experimented (cf. subsection 2.3). Messages were classified according to the type of spatial information they comprised: absolute (ASI), relative (RSI), implicit (ISI), plus absolute-relative (ASI+RSI) and absolute-implicit (ASI+ISI). Subjects' results, grouped according to these categories, are presented in table 4. Subjects' performances in the VP condition are also reported although messages were excluded from target presentations in this condition; they serve as references in the assessment of the effectiveness and efficiency of the user support provided by oral messages in the two other conditions (OP and MP). As the number of scenes varied from one class of messages to the other, error percentages were computed for each condition over 48 samples (ASI), 72 samples (RSI), 24 samples (ISI), 36 samples (ASI+RSI), and 12 samples (ASI+ISI) [10].

For each category of messages, comparisons between results achieved by subjects in the VP, OP and MP situations suggest that absolute and/or relative spatial information improved selection accuracy markedly (*cf.* the ASI, RSI and ASI+RSI types of messages). However, the usefulness of ISI messages seems questionable, at least in the OP condition. Their effectiveness in the MP condition denotes the complexity of the interpretation processes at work in the interpretation of multimodal stimuli.

Average RSI and ISI selection times were much longer in the OP condition (4.03 and 6.12 respectively) than in the other conditions (*i.e.*, 2.86 and 1.84 for VP, 3.12 and 2.1 for MP).

For RSI messages, this effect may be due to the complexity of the visual search strategy induced by relative spatial information when the target is unknown visually. In such cases, the search strategy is likely to include two successive steps: first, localization of the reference object, then exploration of its vicinity in search of the target (Gramopadhye and Madhani, 2001). Each step being longer than the search for a visually known target, the resulting global selection time is necessarily much longer than in the other conditions where the target is visually known.

This interpretation may also explain why RSI messages did not affect selection times in the MP condition noticeably. The target being in the vicinity of the reference object and having been viewed previously, it can be recognized through peripheral vision, so that one eye fixation only is required for locating both the reference object and the target (*cf.* Van Diepen, Wampers, d'Ydewall, 1998).

However, it is also possible that, in the MP condition, subjects tended to adopt a simpler search strategy based exclusively on the available visual information (hence comprising a single visual search step) whenever the oral message induced a complex slow selection strategy. This second interpretation has the advantage to explain why both RSI and ISI messages exerted no perceptible influence on selection times in the MP condition.

As for ISI messages, their exploitation involves complex cognitive processes which may slow down target selection in the OP condition. The poor results observed in the OP condition may also suggest that the usefulness of ISI messages is intrinsically limited.

To sum up, whereas the inclusion of any category of verbal spatial information in multimodal target presentations seems worthwhile, absolute spatial information should be preferred over other information types in the design of oral target presentations, in order to improve both selection times and accuracy.

---

[10] That is 192 samples instead of 6 x 34 = 204 samples (*cf.* the two scenes which were excluded). Two additional scenes (6 x 2 samples) were not taken into account because the corresponding messages did not include any spatial information, the visual salience of the target making such information superfluous.

*Table 4: Results per target presentation mode and type of verbal message.*
*Best results in the OP condition are in bold type, and lowest ones are underscored.*
*Percentages were computed over the total number of samples per message type.*
*(Results in the VP condition are reported, VP being used here as the reference condition).*

| **VP** | **Condition** | **(reference)** | |
|---|---|---|---|
| *Scenes grouped per message type* | *Percentage of errors* | *Average selection time (sec.)* | *Standard deviation (sec.)* |
| ASI | 10.4 | 2.87 | 1.19 |
| RSI | 26.4 | 2.86 | 2.14 |
| ISI | 4.17 | 1.84 | 0.57 |
| ASI+RSI | 13.89 | 3.57 | 1.99 |
| ASI+ISI | 8.33 | 3.54 | 0.98 |
| **0P** | **Condition** | | |
| ASI | **0** | 2.91 | 3.41 |
| RSI | **8.33** | <u>4.03</u> | 5.94 |
| ISI | <u>16.67</u> | 6.12 | 3.78 |
| ASI+RSI | **5.56** | 3.82 | 3.78 |
| ASI+ISI | <u>16.67</u> | <u>5.19</u> | 3.37 |
| **MP** | **Condition** | | |
| ASI | 4.17 | 2.42 | 1.41 |
| RSI | 8.33 | 3.12 | 2.43 |
| ISI | 0 | 2.1 | 1.06 |
| ASI+RSI | 0 | 2.82 | 1.84 |
| ASI+ISI | 0 | 2.98 | 2.53 |

However, these observations, which refine hypothesis C (*cf.* subsection 2.1), should be viewed as working hypotheses rather than reliable, or even tentative, conclusions. Their appropriateness has to be assessed through careful systematic experimentation on a large scale. Analysis of eye-movements (by means of an eye-tracker) would provide invaluable information on visual search strategies, especially on the exact influence of the information content of oral messages on these strategies.

*3.2. Qualitative analyses*

Qualitative analyses were focused on the subjects' errors exclusively, with a view to:
- getting a better understanding of the contribution of verbal messages to assisting users in visual search tasks, and
- obtaining useful knowledge for improving message design.

Analyses use the detailed characterizations of scenes and messages listed in subsection 2.6, as well as the subjects' subjective ratings of the difficulty of the prescribed visual search tasks (*cf.* the post-session questionnaires mentioned in subsection 2.5).

Scenes were filtered so that, in each condition, only the scenes which had occasioned more than one error were considered, on the basis of the following assumption:

> For a given scene in a given condition, the reasons for the failure of one single subject are more likely to originate from the subject's capabilities than from the scene characteristics or the message information content.

*3.2.1 Visual condition*

The main plausible factors at the origin of selection errors observed in the VP condition are presented next. Percentages represent:
- the number of errors which a given characteristic of the scene may explain, by itself or in conjunction with other factors;
- computed over the total number of filtered errors (*i.e.*, 26).

Factors are listed in decreasing order of the percentages of errors they contribute to account for:
- Concerning targets:
  lack of salience (85%), eccentric position in the scene (69%), possible confusions with other objects (69%), unfamiliarity (50%).
- Concerning scenes:
  crowded (69%), unstructured (46%), figuring geometric forms (42%).

This analysis of subjects' errors in the VP condition will be used as a reference in the next subsection which is focused on errors in the OP and MP conditions.

*3.2.2. Oral and multimodal conditions*

Five scenes in the OP condition and only two scenes in the MP condition occasioned more than one error, against eight in the VP condition.
In addition, 24 errors in the VP condition were "corrected" in the OP condition, so that seven out of the eight scenes occasioning more than one error in the VP condition yielded error-free results in the OP condition.
These comparisons bring out the usefulness of oral messages for improving target selection accuracy.

However, four scenes yielding error-free results in the VP and MP conditions occasioned ten out of the twelve [11] filtered errors observed in the OP condition [12]. Therefore, it is likely that the main factor at the origin of these errors is the poor quality of the information content of the oral messages paired off with these scenes.

The analysis of these four messages, together with the information provided by questionnaires and debriefings, support this conclusion. Four errors were motivated by an ISI message which referred to knowledge unfamiliar to the majority of subjects. A too complex ASI+ISI message (structure and length) referring to knowledge some subjects were unfamiliar with may account for two other errors. As for the two remaining pairs of errors, they may be reliably ascribed to the use, in both verbal target designations, of technical substantives the exact meanings of which were unfamiliar to some subjects.

The fact that none of these errors occurred in the MP condition, together with the fact that two scenes only occasioned the four [13] filtered errors observed in this condition, illustrates the advantages of combining visual and verbal information in target presentations.

Two errors occurred on a "difficult" scene which occasioned six errors in the VP condition (crowded scene, and non salient unfamiliar target easy to confound with other objects), and two errors in the OP condition (use of technical vocabulary). The other two errors were occasioned by a scene which was processed successfully by all subjects in the OP condition, but occasioned three errors in the VP condition. This may hint that the processing of multimodal incoming information is controlled or influenced by visual perception strategies rather than by high level cognitive processes.

In short, the qualitative analysis of errors confirms the usefulness of oral messages for improving the accuracy of visual target identification, provided that:
− messages are short, their syntactical structure straightforward, the vocabulary familiar to users, and
− above all, their information content is appropriate.

*3.2.3. Subjects' subjective judgements*

Subjects expressed positive judgments on the contribution of oral messages to facilitating visual search tasks in the post-experiment questionnaires. The MP condition achieved the highest rate of subjective satisfaction as shown in table 5. The majority of subjects (72%) rated the execution of visual search tasks in the MP condition as "very easy", whereas a minority of 22% only applied this rating to the OP and VP conditions. The VP condition got the lowest ratings.

In addition, the MP condition, compared to the VP and OP conditions, was judged most efficient, in terms of rapidity, by fourteen subjects.

---

[11] out of 14.

[12] Error patterns for these images were as follows: 4, 2, 2, 2.

[13] out of 8.

Thus, the majority of subjects considered implicitly that oral messages had helped them to achieve the prescribed search tasks, especially when verbal information was associated with a visual presentation of the target.

Finally, the majority of subjects (66%) preferred the MP condition to the others.

*Table 5: Subjects' judgments (on task difficulty) and preferences.*
*Percentages were computed over the total number of subjects.*
*Highest values in each line are in bold type.*

| Condition | "Very easy" | "Easy" | "Difficult" | "Very difficult" |
|---|---|---|---|---|
| VP | 22% | 28% | **39%** | 11% |
| OP | 22% | **61%** | 17% | 0% |
| MP | **72%** | 17% | 11% | 0% |

| Top-ranking preference | VP | OP | MP |
|---|---|---|---|
| | 17% | 17% | **66%** |

Nevertheless, three subjects complained of the content or wording of some oral messages. Criticisms concern, for one subject, messages without spatial information (two messages), for the second one, the use of a somewhat technical word outside his vocabulary (one message), and for the last one, the length and complexity of RSI messages. In addition, a fourth subject considered oral messages were not useful in the MP condition.

To conclude, the MP condition came first in the subjective judgments of most subjects, as regards both the utility and usability of oral messages. These results are encouraging in view of the numerous potential interactive applications involving visual search tasks.

However, voluntary participants in experimental evaluations of novel artifacts or techniques are prone to judge them positively (experimental bias), especially if the evaluation consists in a single session like ours. Their judgments may evolve under the influence of experience. Further usability studies are needed, in particular for assessing how future users will appraise the support provided by oral messages after extensive practice in real contexts of use.

## 4. CONCLUSION

A preliminary experimental study has been presented, which aims at eliciting the contribution of oral messages to facilitating visual search tasks on crowded visual displays.

Results of quantitative and qualitative analyses suggest that appropriate verbal messages can improve both search accuracy and selection times. In particular, multimodal messages including absolute spatial oral information on the target location in the visual scene, together with a visual presentation of the isolated target, are most effective.

This type of messages also got the highest subjective satisfaction ratings from most subjects. Subjects' acceptance, even in an experimental environment, is a valuable asset.

Numerous potential applications exist. Facilitating visual search could indeed improve the efficiency and usability of present human-computer interaction sensibly, as direct manipulation of GUIs is the prevalent user interface design paradigm for the moment.

However, these results are only tentative, by reason of the relatively small number of subjects involved in the experiment (18), the limited number of scenes they had to process (12 *per* condition), and the coarseness of the measures used which were restricted to search accuracy and selection times (including search, identification, and selection of the target). In addition, qualitative analyses (focused on subjects' errors) suggest the possible influence, on subjects' results, of factors that were not systematically taken into account in the design of our experimental protocol. Our current short term research directions are based on these observations.

We are currently planning a series of experimental studies focused on in-depth investigation of the possible influence, on search accuracy and selection times, of the visual characteristics of scenes and targets, namely scene structure or target position. In particular, structure seems a factor capable of facilitating the exploration of class 1 scenes significantly (see Cribbin and Chen, 2001a), whereas the efficiency of relative spatial information can be influenced by the proximity of the target to the salient object (Gramopadhye and Madhani, 2001).

Each of these experiments will be designed along the same lines as the one presented here, and implemented using a similar experimental protocol. However, it will address a few specific related issues and involve a large number of participants, in order to make it possible to refine and enrich the results of the initial study presented here.

In addition, to achieve sound meaningful interpretations of future quantitative experimental results, we shall compare them with qualitative eye-tracking data obtained from a carefully selected sample of participants.

We believe that a better understanding of visual search strategies, of their inter-individual diversity (Cribbin and Chen, 2001b), their sensitivity to the visual characteristics of the scenes displayed, and their evolution under the influence of oral support, will prove useful for improving the design of oral user support in visual search tasks. In particular, such knowledge could help designers to tailor the information content and wording of oral messages to individual search strategies and visual scene characteristics.

A further step may be to compare, in terms of effectiveness and usability, oral support to visual search with various visual enhancements of targets.